\documentclass[11pt,a4paper]{article}
\usepackage[T1]{fontenc}
\usepackage[utf8]{inputenc}
\usepackage{authblk}

\usepackage{geometry}
\usepackage{amsfonts,amsmath,amssymb,amsthm,nicefrac}
\usepackage{ulem}
\usepackage{cancel}
\usepackage{yhmath}
\usepackage{arcs}
\usepackage{graphicx}
\usepackage{url}
\title{\bf On moduli space of the Wigner quasiprobability distributions for $N$\--dimensional quantum systems}
\author[1]{Vahagn Abgaryan\thanks{vahagnab@googlemail.com}}
\author[1,2,3]{Arsen Khvedelidze\thanks{akhved@jinr.ru}}
\author[1]{Astghik Torosyan\thanks{astghik@jinr.ru}}
\affil[1]{\small{Laboratory of Information Technologies, Joint Institute for Nuclear Research, Dubna, Russia}}
\affil[2]{\small{Institute of Quantum Physics and Engineering Technologies, Georgian Technical University, Tbilisi, Georgia}}
\affil[3]{\small{A. Razmadze Mathematical Institute, Iv.Javakhishvili Tbilisi State University, Tbilisi, Georgia}}

\date{ }

\begin{document}
\maketitle

\begin{abstract}
A mapping between operators on the Hilbert space of 
$N$\--dimensional quantum system and the Wigner quasiprobability distributions defined on the symplectic flag manifold is discussed. The Wigner quasiprobability distribution is constructed as a dual pairing between the density matrix and the Stratonovich-Weyl kernel. It is shown  that the moduli space of the Stratonovich-Weyl kernel is given  by an intersection of the coadjoint orbit space of the $SU(N)$ group and a unit $(N-2)$\--dimensional sphere. The general consideration is exemplified by a detailed description of the moduli space of 2, 3 and 4-dimensional systems.   
\end{abstract}

\section{Introduction}

According to the postulates of the quantum theory, the fundamental description of a physical system is provided by the density operator~\cite{vonNeumann1932}
\begin{equation}
\varrho =\sum_{k}p_k |\psi_k\rangle\langle \psi_k|\,,
\end{equation}
which represents  the quantum statistical ensemble $\{p_k, |\psi_k\rangle \}\,, $ i.e., a set consisting of  vectors  
$|\psi_k\rangle \in \mathcal{H}$ of the Hilbert space $\mathcal{H}$ and their probabilities $p_k $  with  a sum equal to one,   $\sum_k p_k= 1\,. $
The density operator $\varrho$ determines the expectation value $ \mathbb{E}(\hat{{A}})$ of a Hermitian operator $\hat{A}$ acting on $\mathcal{H}\,,$ 
\begin{equation}\label{eq:quantObser}
 \mathbb{E}(\hat{{A}}) =\mbox{Tr}\left[\hat{{A}}\varrho \right], \quad\mbox{with}\quad 
 \mbox{Tr}\left[\varrho\right] =1\,.
 \end{equation}
The latter  is assigned  to a physical observable  associated with  the operator $\hat{A}$.
On the other hand, an ensemble of a classical mechanical system is characterized by a  probability distribution function  $\rho(q,p)\,,$ i.e., the density of the probability to find the system in a state localized in the vicinity of  a phase space point with coordinates $q$ and $p$. Correspondingly, the statistical average, i.e., the expectation value $\mathbb{E}(A)$ of a physical quantity described by the  
function $A(q,p)$ on a phase space is given by the following convolution: 
\begin{equation}\label{eq:classObser}
\mathbb{E}(A)=\int\mathrm{d}\Omega\,A(q,p)\,\rho(q,p), \quad\mbox{with}\quad \int\mathrm{d}\Omega\,\rho(q,p)= 1\,,
\end{equation}
where $\mathrm{d}\Omega $ denotes the normalized volume form of a classical phase space.

Aiming to collate two representations of observables, the classical  (\ref{eq:classObser}) and the quantum (\ref{eq:quantObser}),  the 
so-called Weyl–Wigner invertible mapping between Hilbert space operators and functions on  a phase space has been introduced in  the early stages of the development of quantum mechanics \cite{Weyl1928}-\cite{Moyal1949}. 
The primary elements  of this 
map are two notions: the \textit{symbol of operator}, i.e., a  function $A_W(q,p)$ corresponding to the operator  $A$\,, and the \textit{quasi-distribution function} $W(q,p)$ defined over a phase space. 
As a result, the quantum analogue of the statistical average  
(\ref{eq:classObser})  reads 
\begin{equation}
\mathbb{E}(\hat{{A}})= \int\mathrm{d}\Omega\,A_W(q,p)\,W(q,p), \quad\mbox{with}\quad \int\mathrm{d}\Omega\,W(q,p)= 1\,.
\end{equation}
However, even a quick-look  at  this attempt to build a bridge between classical and quantum statistical pictures shows a lack of their equivalence.
Indeed, one can point out  the following observations:
\begin{itemize}
\item[-] Because of Heisenberg's uncertainty principle, the function $W(q,p)$ has negative values for certain quantum states. Hence it is not a true probability density and is referred to as quasiprobability distribution.
\item[-] Dirac's quantization rule  based on the 
canonical commutator relations makes 
the interplay between  operators and their symbols highly sophisticated.  Replacement  of canonical variables by their quantum counterparts in expressions of functions over the phase-space  faces an ambiguity  of ordering of the corresponding canonical operators.
\footnote{According to Weyl's rule  of quantization \cite{Weyl1928}, any  classical observable $A(\boldsymbol{p},\boldsymbol{q})$\,, i.e., a function on the phase space $\mathbb{R}^{2n}$ with a standard canonical symplectic structure, is associated with an operator $\hat{A}_\omega$ on the Hilbert space $L^2(\mathbb{R}^n)\,$ constructed as the  
\textit{``Weyl quantum Fourier transform"}: 
\begin{equation}
\label{eq:WeylTransform}
A \mapsto \hat{A}_\omega=\int_{\mathbb{R}^{2n}} \mathrm{d}\Omega(\omega )\, \tilde{A}(\boldsymbol{u}, \boldsymbol{v})\exp{\frac{\imath}{\hbar}\left( \boldsymbol{u} \boldsymbol{\hat{P}}+\boldsymbol{v} \boldsymbol{\hat{Q}} \right)} 
\,,\qquad \mathrm{d}\Omega=\omega(\boldsymbol{u},\boldsymbol{v})\,
\mathrm{d}\boldsymbol{u} \mathrm{d}\boldsymbol{v}\,,
\end{equation}
where $\hat{P}$ and  $\hat{Q}$ are operators on $L^2(\mathbb{R}^n)$ obeying canonical commutator relations,
$\tilde{A}(\boldsymbol{u}, \boldsymbol{v})$ is Fourier transform of ${A}(\boldsymbol{u}, \boldsymbol{v})$\,, and the integration measure $\mathrm{d}\Omega$ is defined by a weight function
$\omega(\boldsymbol{u}, \boldsymbol{v})\,.$ Different choice of  $\omega(\boldsymbol{u}, \boldsymbol{v})\,$  is a source of various orderings of non-commutative  operators $\hat{P}$ and $\hat{Q}$. For example, the  factor 
$\omega(\boldsymbol{u}, \boldsymbol{v})=\exp\left(-\frac{\imath}{2}\,\boldsymbol{u}\boldsymbol{v}\right)$
corresponds to a standard ordering of polynomials in mathematical literature when writing first the position coordinate $Q\,,$ then the momentum $P\,.$ The so-called normal ordering  is related to the weight $\omega(\boldsymbol{u}, \boldsymbol{v})=\exp\left(-\frac{1}{4}\,(\boldsymbol{u}^2 + \boldsymbol{v}^2)\right)\,,$ while 
the original Weyl, or symmetric, order complies with $\omega(\boldsymbol{u}, \boldsymbol{v})=1\,.$
The inverse formula that  maps the operator to its symbol belongs to Wigner \cite{Wigner1932}. For a unit weight factor case, $\omega = 1\,,$ the inverse formula reads:  
\begin{equation}
A(\boldsymbol{u}, \boldsymbol{v })=
\frac{1}{(2\pi\hbar)^n}\mbox{tr}\left[\hat{A}_1
\exp{-\frac{\imath}{\hbar}\left( \boldsymbol{u} \boldsymbol{\hat{P}}+\boldsymbol{v} \boldsymbol{\hat{Q}}\right)}\right]\,.
\end{equation}
A further elaboration of Weyl's quantization scheme leads to the non-commutative formulation of mechanics \cite{Moyal1949} and finally to the development of the so-called deformation quantization, cf. \cite{DitoSternheimer}. }
\end{itemize}
In spite of both flaws, Wigner functions
or other formulated quasiprobability distributions, such as Husimi \cite{Husimi1940} and Glauber-Sudarshan \cite{Sudarshan1963,Glauber1963} representations, remain today an important tool for understanding  of interrelations between quantum and  classical statistical descriptions \cite{HilleryOConnellScullyWigner1984}. Moreover, nowadays one can see a growing interest to phase space formulation of quantum mechanics based on the  method of quasiprobability distributions for finite dimensional systems (see e.g.\cite{RoweSandersGuise}-\cite{KlimovedeGuise2} and references therein).  The latter is coming from needs of diverse applications  in quantum optics \cite{ScullyZubairy1997}  and also in quantum information and communications \cite{BengtssonZyczkowki}.
Such an intense usage of quasi-distributions again raises an issue of understanding of the above mentioned shortcomings.\footnote{History going back to Dirac's idea on negative energies teaches us to pay more attention to a ``nonsense"  of negative probabilities. In this context it is  the best  to afford  the following words by  R.Feynman: \textit{"It is that a situation for which a negative probability is calculated is impossible, not in the sense that the chance for it happening is zero, but rather in the sense  that the assumed conditions of preparation or verification are experimentally unattainable" \cite{Feynman1987}. } 
}  

In the present note,  a  problem of construction of quasiprobability distribution functions for generic $N$\--level systems is studied within a purely algebraic approach. 
The basic  mathematical objects in this approach are: a special unitary 
group $G=SU(N)$\,, its Lie algebra $\mathfrak{g}=\mathfrak{su}(N)$:
\begin{equation}
\mathfrak{su}(N) =\{X \in M(N, \mathbb{C}) \ |\  X=-X^\dagger\,,\quad \mbox{tr}X=0\}\,,
\end{equation}
and its dual space $\mathfrak{g}^*=\mathfrak{su}(N)^* \,.$
~\footnote{
Since $\mathfrak{g}$ is a linear space over the real field $\mathbb{R}$\,, one can define a bilinear map   
\(
\langle .\,,\, . \rangle \, : \mathfrak{g}^*\times\mathfrak{g} \to \mathbb{R}\,,
\)
and identify algebra with its dual.
The conventional inner product on $\mathfrak{g}$\,, 
\begin{equation}\label{eq:DPLie}
\langle A\,,\, B \rangle  : =\mbox{tr}\left(A^\dagger B\right)\,,  \qquad A, B \in \mathfrak{su}(N)\,, 
\end{equation}
enables to set up a duality pairing and to realize an isomorphism between $\mathfrak{g}$ and $\mathfrak{g}^*\,.$
}
It is well known that the  universal covering algebra $\mathfrak{U}(\mathfrak{su}(N))$ of the Lie algebra $\mathfrak{su}(N)$ is an arena of the basic objects of N-level  quantum system. Particularly, a state space $\varrho \in \mathfrak{P}_N\,$ is 
defined  as the space of  \textit{positive semidefinite} 
$N\times N$ Hermitian matrices $H_N$ with a unit trace:
\begin{equation}\label{eq:StateSpace}
    \mathfrak{P}_N=\{ X \in H_N \ |\  X \geq 0\,,  \quad \mbox{tr}\left( X \right) = 1   \}\,.
\end{equation}
Every state described by the  density matrix $\varrho \in \mathfrak{P}_N\,$
is in correspondence with some element of the Lie algebra $\mathfrak{su}(N)$:
\begin{equation}\label{eq:DMsuN}
\varrho = \frac{1}{N}\mathbb{I}_N + \frac{1}{N}\,\imath\, \mathfrak{su}(N)\,.
\end{equation}
In order to build up the  Wigner function, apart from the quantum state space $\mathfrak{P}_N\,,$  the notion of its dual $\mathfrak{P}^\ast_N$ is required. Every point of the dual space determines the Stratonovich-Weyl (SW) kernel
\cite{Stratonovich,BrifMann1999}. As it was shown recently in \cite{KhA2018}, the space $\mathfrak{P}^\ast_N\,$ can be defined as follows: 
\begin{equation}\label{eq:SWspace}
    \mathfrak{P}^\ast_N=\{ X \in H_N \ |\ \mbox{tr}\left( X \right) = 1\,, 
    \quad
   \mbox{tr}\left( X^2 \right) = N 
    \}\,.
\end{equation}
It turns out that the dual pairing (\ref{eq:DPLie}) of a density matrix $\varrho \in \mathfrak{P}_N$ and SW kernel $\Delta(\Omega_N) \in \mathfrak{P}^\ast_N$:
\begin{equation}\label{eq:WignerFunction}
W_\varrho(\Omega_N) = \mbox{tr}\left[\varrho
\,\Delta(\Omega_N)\right]\,
\end{equation}
enables us with the proper Wigner function which  satisfies  all  the Stratonovich-Weyl postulates~\cite{Stratonovich,BrifMann1999}.
Taking into account a unit trace condition, SW kernel $\Delta(\Omega_N)$ can be related to the dual of $\mathfrak{su}(N)$:
\begin{equation}\label{eq:SWsuN}
\Delta(\Omega_N) = \frac{1}{N}\mathbb{I}_N + \kappa\,\frac{1}{N}\,\imath\, \mathfrak{su}(N)^\ast \,,
\end{equation}
where $\kappa =\sqrt{{N(N^2-1)}/{2}}\,$ is a normalization constant.
From representations (\ref{eq:DMsuN}) and (\ref{eq:SWsuN}) it follows that  all nontrivial information comes from pairing  between  traceless parts of  a density matrix and SW  kernel. 
In the subsequent sections, after a short overview of the Stratonovich-Weyl postulates, algebraic and geometric aspects of the dual space $\mathfrak{P}^\ast_N$ are discussed. In particular, we establish interrelation between the Wigner functions and  the coadjoint orbits \cite{Kirillov} $\mathcal{O}_{\boldsymbol{r}}$ of $SU(N)$: 
\begin{equation}
\mathcal{O}_{\boldsymbol{r}} =\{UDU^\dagger\,: U\in SU(N)\}\,, 
\end{equation}
where $\boldsymbol{r}$ denotes $N$-tuple of real numbers 
$\boldsymbol{r}={r_1, r_2, \dots, r_N}$
which are elements of the diagonal matrix 
$D=\mbox{diag}||r_1, r_2, \dots, r_N|| $
ordered as $r_1\geq r_2\geq \dots \geq r_N\,$ 
and summed up to zero, 
$\sum_{i=1}^N r_i=0
\,.$
It is then proved  that 
\begin{equation}\label{eq:WFCoAdj}
 W_\varrho(\Omega_N)-\frac{I}{N}\,:\  \mathfrak{P}_N \times \mathcal{O}_{\boldsymbol{r}}\bigl |_{\sum r_i^2=N/(N-1)}\quad  \to \quad  \mathbb{R}\,.
\end{equation}
Furthermore, in order to describe in unitary invariant way  an ambiguity of the Wigner function, we introduce the \textit{moduli space} $\mathcal{P}_N$ of SW kernel as the following coset:
\begin{equation}\label{eq:ModuliOrbit}
\mathcal{P}_N:= \frac{\mathcal{O}_{\boldsymbol{r}}}{SU(N)}
\Bigg |_{\sum r^2_i={N}/(N-1)}\,.
\end{equation}
The moduli space geometrically represents intersections of the orbit space of the  $SU(N)$ group coadjoint action with an ($N-2$)\--dimensional sphere. Finalizing  our note, we give  few examples of  the moduli space of the Wigner functions for low-level quantum systems, for  
a qubit (N=2), qutrit (N=3) and quatrit (N=4).

%
\section{Constructing the Wigner function}

Below we give a brief summary of the Wigner quasiprobability distribution
construction starting from the  basic Stratonovich-Weyl postulates and 
reformulating them into a set of algebraic constraints on a 
spectrum of SW kernels $\Delta(\Omega_N)\,.$

\noindent{$\bullet$ {\bf The Stratonovich-Weyl principles} $\bullet$}
Following to Brif and Mann \cite{BrifMann1999}, the  postulates known as the Stratonovich-Weyl correspondence can be written as the following constraints on the kernel $\Delta(\Omega_N)$:
\begin{enumerate}
\item {\bf Reconstruction}: a state $\varrho$ is reconstructed from the WF (\ref{eq:WignerFunction}) via the integral over a phase space: 
\begin{equation}\label{eq:DMWigner}
\varrho =\int_{\Omega_N} \mathrm{d}\Omega_N\, \Delta(\Omega_N) W_\varrho(\Omega_N) \,;
\end{equation}	
\item {\bf Hermicity}: 
\begin{equation}
\Delta(\Omega_N)= \Delta(\Omega_N)^\dagger\,;
\end{equation}
\item {\bf Finite Norm}: a state norm is given by the integral of the Wigner distribution: 
\begin{equation}
\mbox{tr}[ \varrho ]= \int_{\Omega_N} 
\mathrm{d}\Omega_N W_\varrho(\Omega_N)\,, 
\qquad
\int_{\Omega_N} \mathrm{d}\Omega_N\,\Delta(\Omega_N) = 1\,;
\end{equation}
\item {\bf Covariance}: the unitary transformations $\varrho^\prime = U(\alpha)\varrho U^\dagger(\alpha)$ induce the kernel change:
\begin{equation}
\Delta(\Omega^\prime_N) =U(\alpha)^\dagger\Delta(\Omega_N)U(\alpha)\,.
\end{equation}
\end{enumerate}

\paragraph{Algebraic master equation for SW kernel}
The above given axioms allow derivation of algebraic equations for SW kernel of $N$\-- level quantum systems. With this goal, following the paper \cite{KhA2018},  we accomplish next steps: 
\begin{itemize}
\item[\bf I.]{\bf Identification of phase-space $\Omega_N$ with complex flag manifold}.\\
Hereinafter, 
a phase-space $\Omega_N$ will be  identified with a complex flag manifold, 
$\Omega_N =\mathbb{F}^N_{d_1,d_2, \dots, d_s}\,.$
The latter emerges as follows: supposing  that a spectrum of SW kernel $\Delta(\Omega_N)\,$ consists of real eigenvalues with the algebraic multiplicity $k_i$, i.e., the isotropy group $H$ of the kernel is
\begin{equation}\label{eq:isotropyH}
\nonumber H={U(k_1)\times U(k_2) \times U(k_{s+1})}\,,
\end{equation}
one can see that the phase space $\Omega_N$ can be realized as a coset space $U(N)/H$, the complex flag manifold
$
\mathbb{F}^N_{d_1,d_2, \dots, d_s}\,,
$
where  $(d_1, d_2, \dots, d_s)$ is a sequence of positive integers with sum
$N $, such that  $k_1=d_1$ and $k_{i+1}=d_{i+1}-d_i$ 
with $d_{s+1}=N\,.$ 
Furthermore, since the flag manifold represents a coadjoin orbit of $SU(N)$\,, its symplectic structure 
is given by the corresponding Kirillov-Kostant-Souriau symplectic 2-form~\cite{Kirillov}.
\item[\bf II.]{\bf Enlarging  of phase-space
$\Omega_N$ to $SU(N)$ group manifold.}\\
Owing to the  unitary symmetry of $N$\--dimensional quantum system, we can relate a measure $\mathrm{d}\Omega_N $ on the symplectic space  $\Omega_N$
with the normalized Haar measure $\mathrm{d}\mu_{SU(N)}$ on the $SU(N)$ group manifold: 
\[
\mathrm{d}\Omega_N = C_N^{-1}{\mathrm{d}\mu_{SU(N)}}/{\mathrm{d}\mu_H}\,.
\]
Here $C_N$ is a real normalization constant, 
$\mathrm{d}\mu_{H}$ is the Haar measure on the isotropy group $H$ induced by the embedding, $H \subset SU(N)\,.$
Noting that the integrand in (\ref{eq:DMWigner}) is a function of the coset variables only, the reconstruction integral can be extended to the whole group  $SU(N)$, 
\begin{equation}
\label{eq:reconstovergroup}
\varrho = Z_N^{-1}\int_{SU(N)} \mathrm{d}\mu_{SU(N)}\, \Delta(\Omega_N) W_\varrho(\Omega_N) \,,
\end{equation}	
where the  normalization constant $Z_N^{-1}= C_N^{-1}/\mbox{vol}(H)\,$ includes the factor $\mbox{vol}(H)$ which is the volume of the isotropy group $H$.
\item[\bf III.]{\bf Derivation of algebraic equations for 
SW kernel.}\\
Relations  (\ref{eq:WignerFunction}) and  (\ref{eq:reconstovergroup}) imply the integral identity
\begin{equation}\label{eq:RhoIdentity}
\varrho=Z_N^{-1} \int_{SU(N)} \mathrm{d}\mu_{SU(N)}\,  \Delta(\Omega_N)\,
\mbox{tr}\left[\varrho\Delta(\Omega_N)\right]\,.
\end{equation}
Substituting the singular value  decomposition for SW kernel into (\ref{eq:RhoIdentity}) 
and evaluating  the integral using the Weingarten formula~\cite{Weingarten,Colins2003,ColinsSniady2006}:
\begin{eqnarray}
\nonumber\int{d\mu}U_{i_1 j_1}U_{i_2 j_2}\bar{U}_{k_1 l_1}\bar{U}_{k_2 l_2} 
=\frac{1}{N^2-1}\left(\delta_{i_1 k_1}\delta_{i_2 k_2}\delta_{j_1 l_1}\delta_{j_2 l_2}+\delta_{i_1 k_2}\delta_{i_2 k_1}\delta_{j_1 l_2}\delta_{j_2 l_1}\right)-\\
\nonumber
\frac{1}{N(N^2-1)}\left(\delta_{i_1 k_1}\delta_{i_2 k_2} \delta_{j_1 l_2}\delta_{j_2 l_1}+\delta_{i_1 k_2} \delta_{i_2 k_1}\delta_{j_1 l_1}\delta_{j_2 l_2}\right),
\end{eqnarray}
we derive the equations: 
\begin{equation}
\left(\mbox{tr}[\Delta(\Omega_N)]\right)^2=Z_N N\,,\quad
\mbox{tr}[\Delta(\Omega_N)^2]=Z_N N^2\,.
\end{equation}
\item[\bf IV.]{\bf Normalization of SW kernel.}\\
The constant $Z_N$ in the equation (\ref{eq:reconstovergroup}) can be determined with the aid of the so-called standardization condition, 
\begin{equation}
Z_N^{-1}\int \mathrm{d}\mu_{SU(N)} W_A(\Omega_N)= \mbox{tr}[A]\,. 
\end{equation}
Fixing the normalization constant $Z_N$\,, we finally arrive at  the  
\textbf{``master equations''} for SW kernel:
\begin{equation}\label{eq:master}
\mbox{tr}\left[\Delta(\Omega_N)\right]=1\,,  \qquad  
\mbox{tr}[\Delta(\Omega_N)^2] = N\,.
\end{equation}

\end{itemize}

%
\section{Moduli space: reckoning up  solutions to the ``master equations'' }

Classifying  solutions to  the master equations (\ref{eq:master}), we arrive at the  notion of a \textit{``moduli space''} as  the space  $\mathcal{P}_N\,,$ points of which are associated with the unitary equivalent admissible SW kernel of $N\--$dimensional quantum system. 
Analysis of eq. (\ref{eq:master}) solutions space displays the following properties  of the moduli space $\mathcal{P}_N$: 
\begin{enumerate}
\item $\mbox{dim}\left(\mathcal{P}_N(\boldsymbol{\nu})\right) = N-2\,, $ i.e., a maximal number of continuous parameters $\boldsymbol{\nu}$ characterizing the solution $\Delta(\Omega_N\,|\, \boldsymbol{\nu})$ is $N-2\,;$
\item geometrically, $\mathcal{P}_N$ is represented as an intersection of an ($N-2$)\--dimensional sphere $\mathbb{S}_{N-2}$ 
with the orbit space $\mathfrak{su}(N)^\ast\slash SU(N)$ of $SU(N)$ action on  a dual space $\mathfrak{su}(N)^\ast$:
\begin{equation}
\mathcal{P}_N \cong \mathbb{S}_{N-2}\,\bigcap\,\frac{\mathfrak{su}(N)^\ast}{SU(N)}\,\,.
\end{equation}
\end{enumerate}

In order to become convinced in  above statements, consider the singular value decomposition of SW kernel and assume that the kernel is generic with all eigenvalues distinct. 
\footnote{In this case, the isotropy group of SW kernel is isomorphic to 
($N-1$)\--dimensional torus $\mathbb{T}^{N-1}=\lbrace g\in SU(N): g\--\mbox{diagonal}\rbrace\,.$}  
Using the orthonormal basis $\{\lambda_1, \lambda_2, \dots,\lambda_{N^2-1}\}$ of 
$\mathfrak{su}(N)$\,, the SVD decomposition reads:   
\begin{equation}\label{eq:SWkernelexp}
 \Delta(\Omega_N|\boldsymbol{\nu})=\frac{1}{N}U(\Omega_N)\left[I+\kappa\sum_{\lambda\in H }\mu_s(\boldsymbol{\nu})\lambda_s\right]U(\Omega_N)^\dagger,
\end{equation}
where $\kappa=\sqrt{{N(N^2-1)}/{2}}$\,, and $ H$ is the Cartan subalgebra $H \in \mathfrak{su}(N)\,.$

From the master equation (\ref{eq:master}) it follows that the  coefficients $\mu_s(\boldsymbol{\nu})$ 
live on an ($N-2$)\--dimensional sphere $\mathbb{S}_{N-2}(1)$ 
of radius one:
 \begin{equation}\label{eq:moduliWF}
 \sum_{s=2}^{N}\mu^2_{s^{2}-1}(\boldsymbol{\nu}) = 1\,.  
\end{equation}
A generic SW kernel can be parameterized by $N-2$ spherical angles. The parameter $(\boldsymbol{\nu})$ introduced in order to label members of the family of the Wigner functions can be associated with a point on $\mathbb{S}_{N-2}(1)\,$. More precisely, a one-to-one correspondence between points on this sphere and unitary non-equivalent SW kernels occurs only for a certain subspace of $\mathbb{S}_{N-2}(1)$\,.
This subspace    
$\mathcal{P}_N({\boldsymbol{\nu}}) \subset \mathbb{S}_{N-2}(1)\,$  represents the moduli space of SW kernel.  Its geometry is  determined by the $\Delta(\Omega_N\,|\,\boldsymbol{\nu})$ eigenvalues ordering.
The chosen descending order of the eigenvalues restricts the range  of spherical angles parameterizing (\ref{eq:moduliWF}) and cuts out the moduli space of $\Delta(\Omega_N\,|\,\boldsymbol{\nu})$ in the form of a spherical polyhedron. Details of SW kernels parametrization in terms of spherical angles are given in the  
Appendix \ref{AppendixA}.
\section{The Wigner function as dual pairing between $\varrho$ and $\Delta$} 

As soon as the space of all possible SW kernels is known, the construction of the Wigner function reduces  to a computation  of pairing (\ref{eq:WignerFunction}).
Using the $\mathfrak{su}(N)$ expansions (\ref{eq:DMsuN})  for  a density matrix $\varrho_\xi$ of $N$\--level system characterized by   ($N^2-1$)\--dimensional Bloch vector
$\boldsymbol{\xi}$\,,
\begin{equation}
\label{eq:rhoN}
\nonumber \varrho_{\xi}=\frac{1}{N}\left(I+\sqrt{\frac{N\left(N-1\right)}{2}}\left(\boldsymbol{\xi},\boldsymbol{\lambda}\right)\right)\,, 
\end{equation}
and SW kernel decomposition (\ref{eq:SWkernelexp}), 
we arrive at the general representation for the WF:   
\begin{equation}\label{eq:WFCartan}
W^{(\boldsymbol{\nu})}_{\boldsymbol{\xi}} (\theta_1,\theta_2, \dots,  \theta_d)=\frac{1}{N}\left[1 + \frac{N^2-1}{\sqrt{N+1}}\,(\boldsymbol{n}, \boldsymbol{\xi})\right]\,,
\end{equation}
where ($N^2-1$)\--dimensional vector $\boldsymbol{n}$ is given by a linear combination of $N-1$ orthonormal  vectors $\boldsymbol{n}^{(s^{2}-1)}$ with coefficients 
$\mu_{s^2-1}(\boldsymbol{\nu})\,,$  $s=2,3,\dots,  N\,,$
\begin{eqnarray}
\nonumber \boldsymbol{n} = \mu_3 
\boldsymbol{n}^{(3)} + \mu_8 \boldsymbol{n}^{(8)}+\dots+ 
\mu_{N^{2}-1}\boldsymbol{n}^{(N^{2}-1)}\,.  
\end{eqnarray}
The vectors $\boldsymbol{n}^{(s^{2}-1)}$ are determined by the Cartan subalgebra $\lambda_{s^2-1} \in H$:
\begin{equation}
\nonumber \boldsymbol{n}^{(s^2-1)}_\mu = \frac{1}{2}\,\mbox{tr}\left( U\lambda_{s^2-1}U^\dagger\lambda_\mu \right)\,, \quad
 s=2,3,\dots,N\,.
\end{equation}
As it was mentioned in the Introduction, the number of the symplectic coordinates  $\vartheta_1, \vartheta_2, \dots , \vartheta_d$  of the Wigner function depends on the  isotropy group of SW kernel (cf. details in \cite{KhA2018}).


\section{Examples}

Below we present an explicit parametrization for a moduli space 
of a few  low-dimensional quantum systems, including a single qubit, qutrit and quatrit.

\subsection{The moduli space of a single qubit SW kernel}

For a 2-level quantum system, a qubit,   the master equations
(\ref{eq:master}) determine the spectrum (up to permutation) of  2-dimensional SW kernel uniquely:
\begin{equation}\label{eq:Deltaqubit}
\Delta^{(2)}(\Omega_2)=
\frac{1}{2}
U(\Omega_2)
\begin{pmatrix}
1+\sqrt{3} & 0 \\
0 & 1-\sqrt{3}
\end{pmatrix}
U(\Omega_2)^\dagger\,,
\end{equation}
with $U(\Omega_2)  \in SU(2)/ U(1)\,.$ 
Its connection  to the structure of the coadjoint orbits of $SU(2)$ is straightforward. 
There are two types of the  coadjoint orbits of $SU(2)$:  
\begin{enumerate}
\item 2-dimensional regular orbits $\mathcal{O}_{\boldsymbol{r}}$\,,
\[
\mathcal{O}_{\{r, - r\}} =\left\lbrace\,  U
\begin{pmatrix}
r & 0 \\
0 & -r
\end{pmatrix}U^\dagger\,, \, U\in SU(2)\,
\right\rbrace\,,
\]
defined for an ordered 2-tuple, $\boldsymbol{r}=\{r,-r\}\,,$  $r >0\,.$ They are isomorphic to a 2-dimensional sphere $\mathbb{S}_2(r)$
with the radius given by the value of the $SU(2)$ invariant:
\begin{equation}
r^2=-\det\left(\mathcal{O}_{\boldsymbol{r}}\right)\,;
\end{equation}
\item zero-dimensional orbit,  point $ r=0\,.$
\end{enumerate}
Identifying these orbits with the traceless part of SW kernel $\Delta^{(2)}-\frac{1}{2}\mathbb{I}$ and taking into account the expression (\ref{eq:Deltaqubit}), we get convinced that 
\[
r^2=
\frac{4}{3}\,\mbox{tr}
\left[\left(\Delta^{(2)}-\frac{1}{2}\,\mathbb{I}\right)^2\right]
=2\,.
\]
Thus,  the  moduli space of SW kernel of a qubit represents the single point, $r^2=2\,,$ from the set of equivalence  classes of the regular $SU(2)$ orbits,    $[\mathcal{O}_{\boldsymbol{r}}] \cong \mathfrak{su}(2)/U(1)$\,.

\begin{figure} 
\begin{center}
{\includegraphics[width=0.4\linewidth]{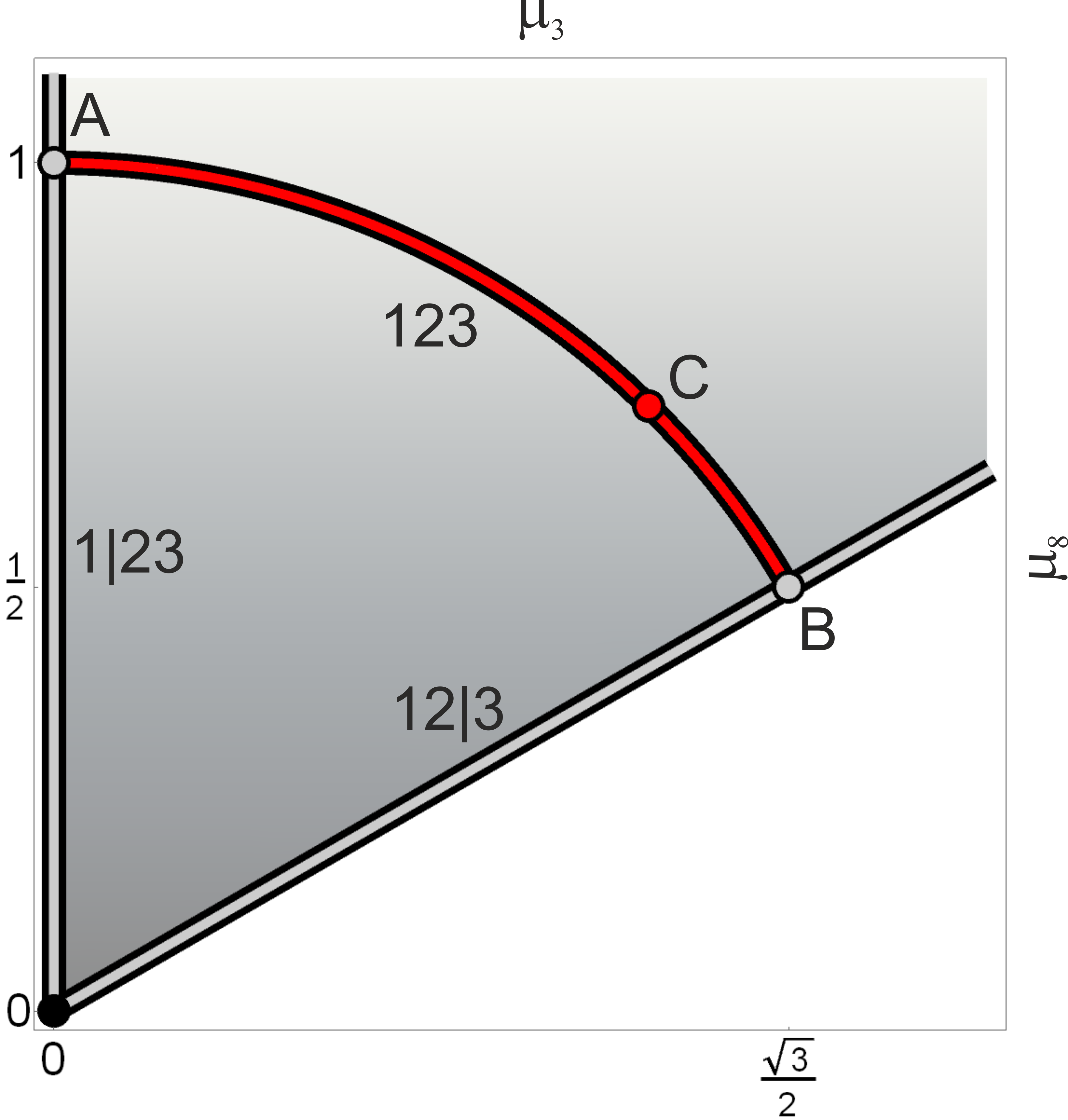}}
\caption{The cone representing the orbit space of $SU(3)$. The interior of the cone represents $\dim{\mathcal{O}}=4$ orbits.  The apex corresponds to a zero-dimensional orbit, while other points on the ordinate $(\mu_8=0)$ and on the 
positive ray $\mu_8=\mu_3/\sqrt{3}$ also determine $\dim{\mathcal{O}}=4$ orbits.
The intersection of the cone with a unit circle gives an arc which is the moduli space of a qutrit SW kernel. The point $C$ with    
$
\cos(\zeta_{C}) = (-1+3\sqrt{5})/{8}\,
$
describes the singular SW kernel. 
}
\label{Fig:Qutrit_Kernel_Mod}
\end{center}
\end{figure}

\subsection{The moduli space of a single qutrit SW kernel}

The master equations (\ref{eq:master})  determine two lowest-degree polynomial $SU(N)$ invariants of SW kernel $\Delta(\Omega_3)\,,$ the linear and the quadratic ones. 
For the case of a 3-dimensional quantum system, a qutrit, 
the  third algebraically independent  polynomial $SU(3)$ invariant remains  unfixed, thus allowing a one-parametric family of a qutrit SW kernels to  exist.

Following the normalization convention (\ref{eq:SWkernelexp}), let us write down the SVD decomposition of a qutrit SW kernel in the following form:
\begin{equation}\label{eq:QutrutSWEigen}
\Delta(\Omega_3)=U(\Omega_3)PU(\Omega_3)^\dagger= U(\Omega_3)\left[\frac{1}{3}\mathbb{I}+
\frac{2}{\sqrt{3}}\,
\begin{pmatrix}
r_1 & 0   & 0 \\
0   & r_2 & 0 \\
0   & 0   &r_3
\end{pmatrix}
\right]
U(\Omega_3)^\dagger\,,
\end{equation}
where $U(\Omega_3) \in SU(3)$\,, and a 3-tuple $\boldsymbol{r}=\{r_1, r_2,  r_3 \} $ parameterizes a traceless diagonal part of the SVD decomposition of SW kernel, \(
r_1+r_2+r_3=0\,.
\)
Expanding  $P$ over the Gell-Mann basis elements $\lambda_3=\mbox{diag}||1, -1, 0||$ and $\lambda_8=\frac{1}{\sqrt{3}}\mbox{diag}||1,1,-2||$ of the $SU(3)$ Cartan subalgebra, 
\begin{equation}\label{eq:diagonalSW}
P=\frac{1}{3}\mathbb{I}+
\frac{2}{\sqrt{3}}\left(\mu_3\lambda_3 +\mu_8\lambda_8\right)\,,
\end{equation}
we find: 
\begin{eqnarray}\label{eq:EigeMu}
r_1=\mu_3+\frac{1}{\sqrt{3}}\mu_8\,,\quad \
r_2=-\mu_3+\frac{1}{\sqrt{3}}\mu_8\,,\quad\
r_3=-\frac{2}{\sqrt{3}}\mu_8\,.
\end{eqnarray}
From these relations it follows that the chosen decreasing order of parameters $r_1\geq r_2\geq r_3\,$  determines on the $(\mu_3, \mu_8)$\--plane the 2-dimensional polyhedral cone $C_2({\pi}/{3})$ with the apex angle $\pi/3$:
\begin{equation}
C_2({\pi}/{3})=\left\lbrace \boldsymbol{x}  \in \mathbb{R}^2\,\ \bigg|\ \, 
\begin{pmatrix}
1 & 0    \\
\frac{-1}{\sqrt{3}}   & 1 
\end{pmatrix} 
\begin{pmatrix}
x_1  \\
x_2 
\end{pmatrix} 
\geq 0\,
 \right\rbrace\,.
\end{equation}
\noindent{$\bullet$ {\bf The SU(3) orbits} $\bullet$}
The cone $C_2({\pi}/{3})$ represents the orbit space of $SU(3)$ group action on $\mathfrak{su}(3)$ algebra.  
Next we identify this algebra times $\imath $ with the traceless part of $\Delta(\Omega_3)$  and  classify SW kernel in accordance to the corresponding coadjoint orbits. In order to realize this program, let us  
consider the tangent space to the $SU(3)$ orbits. It is spanned by the linearly independent vectors built of the commutators: 
\(
t_k = [\lambda_k,\Delta], \ \lambda_k \in \mathfrak{su}(3)\,.
\)
The number of independent vectors $t_k$ determines the dimensionality  
of the orbits via the rank of the  $ 8\times 8$ Gram matrix:
\begin{equation}\label{eq:Gram}
\mathcal{G}_{kl}(\Delta^{(3)}) = \frac{1}{2}\, \mbox{tr}\,(t_k t_l)\,, \qquad
k,l =1, 2, \dots, 8\,.
\end{equation}
Since the rank of the Gram matrix (\ref{eq:Gram}) is $SU(3)$ invariant, one can calculate it for the diagonal representative of SW kernel  (\ref{eq:diagonalSW}). The straightforward  computations give 
\begin{equation}
\mathcal{G}(\Delta^{(3)})=\frac{4}{3} 
\mbox{diag}\,||\, g_1\,, g_1\,, 0\,, g_2\,, g_2\,, g_3\,, g_3\,, 0\,||\,,
\end{equation}
where $g_1 =4\mu_3^2$\,, \quad $g_2 =\frac{1}{\sqrt{3}}(\mu_3+\sqrt{3}\mu_8)^2$\,, \quad 
$g_3 =\frac{1}{\sqrt{3}}(\mu_3-\sqrt{3}\mu_8)^2$\,.
From these expressions it follows that there are three types of $SU(3)$ orbits which can be  classified according to  their symmetry and dimensions:  
\begin{enumerate}
\item $\underline{\dim(\mathcal{O}_{\boldsymbol{r}})=6\,.}$
These \textit{regular orbits} abbreviated as $\mathcal{O}(123)$ (or simply $123$) are labeled by a 3-tuple $\boldsymbol{r}$ with $r_1 > r_2 > r_3\,$ and have the isotropy group $H_{(123)}$ isomorphic to a 2-dimensional torus, 
$H_{(123)}\cong\mathbb{T}^2$\,. They are in one-to-one correspondence with the interior points of the cone $C_2({\pi}/{3})$ in Fig.\ref{Fig:Qutrit_Kernel_Mod}.
\item $\underline{\dim(\mathcal{O}_{\boldsymbol{r}})=4\,.}$ 
These  \textit{degenerate orbits} represent two subfamilies with degenerate 3-tuples $\boldsymbol{r}$: either  $r_1 = r_2 > r_3\,$ or $r_1 >r_2 = r_3\,.$
Following  V.I.Arnold \cite{Arnold}, we denote them as $1|23$ and $12|3$ respectively. Geometrically, the equivalence class $[\mathcal{O}]$ of degenerate orbits represents the boundary lines in the $SU(3)$ orbit space: 
\begin{eqnarray}
\nonumber
&\mathcal{O}(1|23)\mapsto 1|23:&\{\boldsymbol{x}\in C_2({\pi}/{3}) |\ x_2 = 0\ \}\,,\\
&\mathcal{O}(12|3)\mapsto 12|3:& 
\{\boldsymbol{x}\in C_2({\pi}/{3}) |\ x_2=x_1/\sqrt{3}\ \}\,. 
\nonumber
\end{eqnarray}
Both classes, up to conjugacy in $SU(3)$\, have the same isotropy group:
\begin{equation}
H_{(12|3)} \cong H_{(1|23)}=\left\lbrace \ h \in \left[  
\begin{array}{c|c}
e^{i\alpha}g & 0 \\
\hline
0 & e^{-i\alpha}
\end{array}
\right]\
 \bigg|\ g \in SU(2)\  \right\rbrace\,.
\end{equation}
\item $\underline{\dim(\mathcal{O}_{0})=0\,.}$ One orbit $\mathcal{O}_{0}\,,$
a single point $(0,0)\,$ which is stationary under the $SU(3)$ group action.
\end{enumerate}

\noindent{$\bullet$ {\bf The parametrization of a qutrit SW kernels} $\bullet$}
We are now in a position to describe the moduli space of a qutrit as a certain subspace of the $SU(3)$ orbit space. Indeed, taking into account that the second order master equation (\ref{eq:master}) describes a  circle of radius one centered at the origin of  $(\mu_3, \mu_8)$\--plane,  we convinced that the moduli space of a qutrit SW kernel represents the arc depicted in  Fig. \ref{Fig:Qutrit_Kernel_Mod}.  
More precisely, based on the above classification of the $SU(3)$ orbits,  we treat a qutrit moduli space as the union of two strata:
\begin{itemize}
\item The regular stratum corresponding to the regular $SU(3)$\--orbits. Geometrically it is the arc 
$\wideparen{AB}/\{A,B\}$
with its endpoints $A$ and $B$ excluded. The corresponding Wigner functions have a 6-dimensional support and  1-dimensional  family of SW kernels, the spectrum of which can be written as:
\begin{equation}\label{eq:qutritSWnuparamet}
\mbox{spec}\left(\Delta^{(3)}(\nu)\right)= \left\{\frac{1-\nu+\delta}{2}\,,\,\frac{1-\nu-\delta}{2}\,,\,\nu\right\}\,,
\end{equation}
where $\delta=\sqrt{(1+\nu) (5-3\nu)}\,$ and $\nu \in (-1/3, -1)\,.$ The parameter $\nu$ is related to the apex  angle  $\zeta$ of the cone $C_2(\pi/3)$:
\footnote{The apex angle $\zeta$ determines the value of a 3-rd order polynomial SU(3)-invariant:
\[
\cos(3\zeta)=-\frac{27}{16}\,\det\left(\Delta^{(3)}-\frac{1}{3}\mathbb{I}\right)=-\frac{27}{16}\det\left(\Delta^{(3)}\right)-
\frac{11}{16}\,. 
\]
}
\begin{equation}
\nu=\frac{1}{3}-\frac{4}{3}\cos(\zeta)\,, \quad \zeta \in [0,\  {\pi}/{3}]\,.
\end{equation}

\item The end points $A$ and $B$ of the arc 
$\wideparen{AB}$ correspond to two degenerate SW kernels, with 
$\nu=-1$ and $\nu=-\frac{1}{3}$  respectively,  
\begin{equation}
\nonumber
\mbox{spec}\left(\Delta^{(3)}({-1})\right)=
\left\{1, 1, -1\right\},\quad 
\mbox{spec}\left(\Delta^{(3)}({-\frac{1}{3}})\right)=
\frac{1}{3}\left\{5,-1,-1\right\}\,.
\end{equation}
\end{itemize}
It is necessary to point out that the kernel $\Delta^{(3)}({-1})$ was found by Luis \cite{Luis2008}. 

\noindent{$\bullet$ {\bf The singular SW kernels of qutrit} $\bullet$} 
Apart from the above categorization of SW kernels, we distinguish the \textit{singular kernels} which have at least one zero eigenvalue.
From the expression (\ref{eq:qutritSWnuparamet}) it follows that 
for a qutrit case among three zeros of the determinant 
\(\det(\Delta^{(3)})=\nu(\nu^2-\nu-1)\,\) only one, $\nu=(1-\sqrt{5})/{2}\,,$ 
is admissible:
\footnote{Traces of powers of this \textit{``golden ratio''} kernel are given by the so-called \textit{Lucas numbers}:
\[
\mbox{tr}\left(\Delta_{(103)}\right)^2=3\,, \qquad 
\mbox{tr}\left(\Delta_{(103)}\right)^3=4\,, \qquad \ldots \,, \qquad  
\mbox{tr}\left(\Delta_{(103)}\right)^n= L_n\,.
\]
}
\begin{equation}
\nonumber
\mbox{spec}\left(\Delta_{(103)}\right) =
\left\{\frac{1+\sqrt{5}}{2}\,, \, 0\,, \, \frac{1-\sqrt{5}}{2}\right\}\,.
\end{equation}

%
\section{The moduli space of a single quatrit SW kernel}

The master equations (\ref{eq:master}) for a four-level system, a quatrit,  determine a 2-parametric family of SW kernels. 
We start, similarly  to a qutrit case, with the SVD decomposition of a quatrit SW kernel:
\begin{equation}\label{eq:QuatritSWEigen}
\Delta(\Omega_4)=U(\Omega_4)PU(\Omega_4)^\dagger= U(\Omega_4)\left[\frac{1}{4}\mathbb{I}+
\frac{\sqrt{30}}{4}\,
\begin{pmatrix}
r_1 & 0   & 0 & 0\\
0   & r_2 & 0 &0\\
0   & 0   &r_3 & 0\\
0   & 0   & 0 & r_4
\end{pmatrix}
\right]
U(\Omega_4)^\dagger\,,
\nonumber
\end{equation}
with $U(\Omega_4) \in SU(4)$ and a 4-tuple $\boldsymbol{r}=\{r_1, r_2,  r_3, r_4 \}\,,$ such that  \(r_1+r_2+r_3 +r_4 =0\,.
\)
These  parameters expressions in terms of expansion coefficients of  $P$ over the Gell-Mann basis elements $\lambda_3=\mbox{diag}||1, -1, 0, 0||$, $\lambda_8=\frac{1}{\sqrt{3}}\mbox{diag}||1,1,-2,0||$ and $\lambda_{15}=\frac{1}{\sqrt{3}}\mbox{diag}||1,1,1,-3||$ of the $SU(3)$ Cartan subalgebra, 
\begin{equation}\label{eq:diagonalSW4}
P=\frac{1}{4}\mathbb{I}+
\frac{\sqrt{30}}{4}\left(\mu_3\lambda_3 +\mu_8\lambda_8 +\mu_{15}\lambda_{15}\right)\,,
\end{equation}
read:
\begin{eqnarray}\label{eq:EigeMu4}
&&r_1=\mu_3+\frac{1}{\sqrt{3}}\mu_8+
\frac{1}{\sqrt{6}}\mu_{15}\,,\quad \
r_2=-\mu_3+\frac{1}{\sqrt{3}}\mu_8+
\frac{1}{\sqrt{6}}\mu_{15}\,,\\
&&r_3=-\frac{2}{\sqrt{3}}\mu_8 +
\frac{1}{\sqrt{6}}\mu_{15}\,, \qquad \quad  
r_4=-\frac{3}{\sqrt{6}}\mu_{15}\,.
\end{eqnarray}
Due to the order $r_1\geq r_2\geq r_3\geq r_4$\,,  expansion coefficients $\mu_3, \mu_8$ and $ \mu_{15}$ 
belong  to a  3-dimensional polyhedral cone $C_3\left(\pi/6\right)$ with the apex angle $\pi/6$:  
\begin{equation}
C_3\left({\pi}/{6}\right)=\left\lbrace \boldsymbol{x}  \in \mathbb{R}^3\,\ \bigg|\ \, 
\begin{pmatrix}
1 & 0  & 0  \\
\frac{-1}{\sqrt{3}}   & 1 & 0 \\
0& \frac{-1}{\sqrt{2}}   & 1 
\end{pmatrix} 
\begin{pmatrix}
x_1  \\
x_2 \\
x_3
\end{pmatrix} 
\geq 0\,
 \right\rbrace\,.
\end{equation}

\noindent{$\bullet$ {\bf The SU(4) orbits} $\bullet$} The cone $C_3\left({\pi}/{6}\right)$ represents the  $SU(4)$ orbit space.
The calculated for a diagonal representative $15\times15$ Gram  matrix   
\begin{equation}\label{eq:Gramquatrit}
\mathcal{G}(\Delta^{(4)})=\frac{5}{2}\,\mbox{diag}||g_1\,, g_1\,, 0\,, g_2\,, g_2\,, g_3\,, g_3\,, 0\,, g_4\,, g_4\,, g_5\,, g_5\,, g_6\,, g_6\,, 0||\,,
\end{equation}
where
\begin{eqnarray}
&& g_1 = 3 \mu _3^2\,, \quad 
g_2 = \frac{3}{4} \left(\mu _3+\sqrt{3} \mu _8\right)^2\,, \quad 
g_3 = \frac{3}{4} \left(\mu _3-\sqrt{3} \mu _8\right)^2\,, \nonumber\\
&& g_4 = \frac{1}{8} \left(\sqrt{6} \mu _3+\sqrt{2} \mu _8+4 \mu _{15}\right)^2\,, \quad
g_5 = \frac{1}{8} \left(-\sqrt{6} \mu _3+\sqrt{2} \mu _8+4 \mu _{15}\right)^2\,, \nonumber\\
&& g_6 = \left(\mu _8-{\sqrt{2}}\mu _{15}\right)^2\,.
\nonumber
\end{eqnarray}
Analysis of zeros of the Gram  matrix (\ref{eq:Gramquatrit}) shows the following pattern of the regular  and degenerate $SU(4)$ orbits.    
\begin{itemize}
\item $\underline{\dim(\mathcal{O}_{\boldsymbol{r}})=12\,.}$
The regular orbits have a maximal dimension owing to the smallest isotropy group: $H_{(1234)}=\mathbb{T}^3\in SU(4)\,.$ The equivalent class of the regular orbits  represents an interior of the cone $C_3\left({\pi}/{6}\right)$;
\item The degenerate orbits are divided  into subclasses:
\begin{enumerate}
\item $\underline{\dim(\mathcal{O}_{\boldsymbol{r}})=10\,.}$ 
The equivalence class of these orbits is one of the following faces of the cone  $C_3\left({\pi}/{6}\right)$:
\begin{eqnarray}
&\mathcal{O}(1|234)\mapsto 1|234: &\{\boldsymbol{x}\in C_3\left({\pi}/{6}\right)\ |\ x_1 = 0\  \}\,,\\
&\mathcal{O}(12|34)\mapsto 12|34:& 
\{\boldsymbol{x}\in C_3\left({\pi}/{6}\right)\ |\ x_1=-\sqrt{3}x_2 \}\,, \\
&\mathcal{O}(123|4)\mapsto 123|4:& 
\{\boldsymbol{x}\in C_3\left({\pi}/{6}\right)\ |\ x_2= +\sqrt{2}x_3  \}\,. 
\end{eqnarray}
All the above orbits have the same isotropy group (up to SU(4)\--con\-ju\-gation):

$
H_{(1|234)}=\left\lbrace \ h \in \left[  
\begin{array}{c|c|c}
e^{i\alpha}g & 0 &0\\
\hline
0 & e^{i\beta}&0\\
\hline
0 &0 &e^{i\gamma}
\end{array}
\right]\
 \bigg|\ g \in SU(2)\,,\ \alpha+\beta+\gamma=0\ \right\rbrace\,.
$

The dimension of this stratum is in agreement with the dimension of the corresponding isotropy group, 
$$\dim(\mathcal{O}_{\boldsymbol{r}})= \dim(SU(4))-\dim(H_{\boldsymbol{r}})=15-(3+2)=10\,.$$

\item $\underline{\dim(\mathcal{O}_{\boldsymbol{r}})=8\,.}$ 
The equivalence class of these orbits is the following edge of the cone $C_3\left({\pi}/{6}\right)$:
\begin{eqnarray}
&\mathcal{O}(1|23|4)\mapsto 1|23|4: &\{\boldsymbol{x}\in C_3\left({\pi}/{6}\right)\ |\ x_1 = 0\,,\ x_2=\sqrt{2}x_3\  \}\,.
\end{eqnarray}
The 7-dimensional isotropy group is:
\begin{equation}
H_{(1|23|4)}=\left\lbrace \ h \in \left[  
\begin{array}{c|c}
e^{i\alpha}g & 0 \\
\hline
0 & e^{-i\alpha}g^\prime
\end{array}
\right]\
 \bigg|\ g, g^\prime \in SU(2)\  \right\rbrace\,.
\end{equation}

\item $\underline{\dim(\mathcal{O}_{\boldsymbol{r}})=6\,.}$ 
The equivalence class of these orbits is one of the following edges of the cone $C_3\left({\pi}/{6}\right)$:
\begin{eqnarray}
&\mathcal{O}(1|2|34)\mapsto 1|2|34: &\{\boldsymbol{x}\in C_3\left({\pi}/{6}\right)\ |\ x_1 = 0\,,\ x_2=0\  \}\,,\\
&\mathcal{O}(12|3|4)\mapsto 12|3|4: &\{\boldsymbol{x}\in C_3\left({\pi}/{6}\right)\ |\ x_1 =\sqrt{3}x_2\,,\ x_2=\sqrt{2}x_3\  \}\,.
\end{eqnarray}
Both classes have the same up to conjugacy 9-dimensional isotropy group:
\begin{equation}
H_{(1|2|34)}=\left\lbrace \ h \in \left[  
\begin{array}{c|c}
e^{i\alpha}g & 0 \\
\hline
0 & e^{-i\alpha}
\end{array}
\right]\
 \bigg|\ g \in SU(3)\  \right\rbrace\,.
\end{equation}

\item $\underline{\dim(\mathcal{O}_{\boldsymbol{r}})=0\,.}$The apex of cone  $C_3\left({\pi}/{6}\right)$
with the stability group $SU(4)\,.$ 
\end{enumerate}
\end{itemize}

\begin{figure}[ht]
\centering
\includegraphics[width=0.6\textwidth]{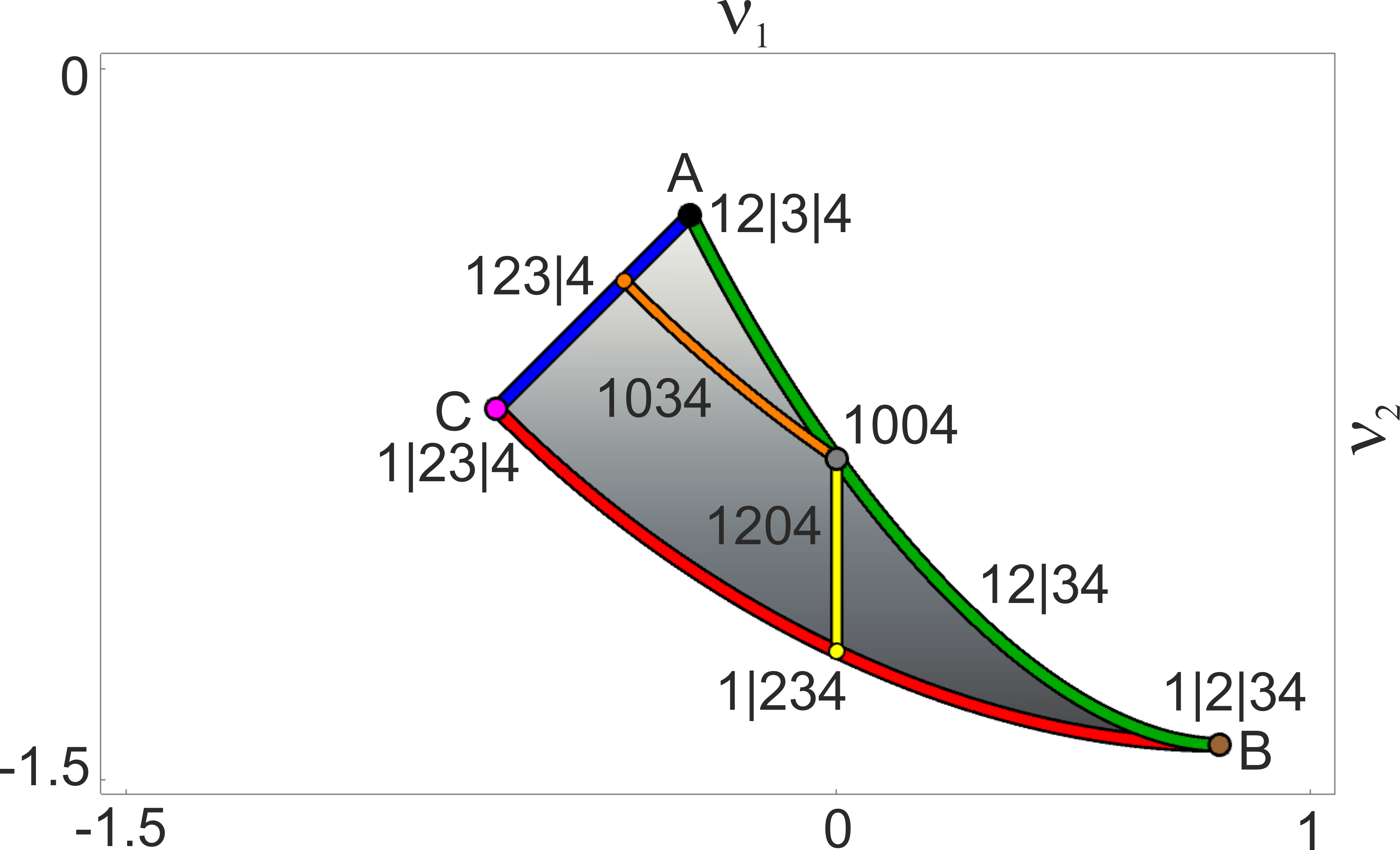}
\caption{Support of SW kernel of a quatrit on $(\nu_1,\nu_2)$\--plane. 
The interior of a curvilinear triangle $ABC$ corresponds to the regular SW kernels.
The boundary lines describe the double degeneracy cases.
The vertexes $A$ and $B$  describe a quatrit kernels with a triple degeneracy,
while the vertex $C$ corresponds to a quatrit kernel with two double degeneracy.}
{\label{Fig:QbQbKernelSimplex}}
\end{figure}

\bigskip

\noindent{$\bullet$ {\bf The parametrization of a quatrit SW kernels} $\bullet$}
Now we are ready to enumerate all SW kernels for a quatrit according to the above given classification of the $SU(4)$ orbits:  
\begin{enumerate}
\item \underline{The regular 2-dimensional family of SW kernels}:
\begin{equation}\label{eq:quatritSWregular}
\mbox{spec}\left(\Delta^{(4)}(\nu_1,\nu_2)\right)= 
\left\{\frac{1-\nu_1-\nu_2+\delta}{2}\,, \frac{1-\nu_1-\nu_2-\delta}{2}\,,  \nu_1\,, \nu_2 \right\}\,,
\end{equation}
where $\delta=\sqrt{7+2\nu_{1}-3\nu_{1}^2+2\nu_{2}-2\nu_{1}\nu_{2}-3\nu_{2}^2}\,.$
\medskip

\item \underline{The degenerate 1-dimensional family of SW kernels}: 
\vspace{0.3cm}

\begin{enumerate}
\item A family of SW kernels of 1|234 type:  
\hfill(red in color)
\begin{equation}
\mbox{spec}\left(\Delta_{(1|234)}\right)=
\left\{\frac{1-\nu}{3}+\frac{1}{6}\delta_1\,, \frac{1-\nu}{3}+
\frac{1}{6}\delta_1\,, \nu\,, 
\frac{1-\nu-\delta_1}{3} \right\}\,,
\end{equation}
where $\delta_1=\sqrt{22+4\nu-8\nu^2}$ and $\nu\in \big (\frac{1}{4}\left(1-\sqrt{15}\right),\frac{1}{4} \left(1+\sqrt{5}\right)\big)$\,;

\item A family of SW kernels of 12|34 type:   
\hfill(green in color)
\begin{equation}\label{eq:quatritSW12|34}
\mbox{spec}\left(\Delta_{(12|34)}\right)= 
\left\{\frac{1-2\nu+\delta_2}{2}\,, \nu\,, \nu\,,  \frac{1-2\nu-\delta_2}{2} \right\}\,,
\end{equation}
where $\delta_2=\sqrt{7+4\nu-8\nu^2}$ and 
$\nu \in \big (\frac{1}{4} \left(1-\sqrt{5}\right),\frac{1}{4} \left(1+\sqrt{5}\right)\big)$\,;

\item A family of SW kernels of 123|4 type:    
\hfill(blue in color)
\begin{equation}\label{eq:quatritSW123|4}
\mbox{spec}\left(\Delta_{(123|4)}\right)= 
\left\{\frac{1-2\nu+\delta_2}{2}\,, \frac{1-2\nu-\delta_2}{2}\,, \nu\,, \nu \right\}\,,
\end{equation}
where  $\nu \in \big (\frac{1}{4} \left(1-\sqrt{15}\right),\frac{1}{4} \left(1-\sqrt{5}\right)\big)$\,.
\end{enumerate}
\medskip

\item \underline{SW kernels with a triple degeneracy}: 
\vspace{0.3cm}

\begin{enumerate}
\item SW kernel of 1|2|34  type: 
\hfill (brown point)
\begin{equation}\label{eq:quatritSW1|2|34}
\mbox{spec}\left(\Delta_{(1|2|34)}\right)= 
\left\{\frac{1+\sqrt{5}}{4}\,, \frac{1+\sqrt{5}}{4}\,, \frac{1+\sqrt{5}}{4}\,, \frac{1-3\sqrt{5}}{4} \right\}\,;
\end{equation}
\item SW kernel of 12|3|4 type: 
\hfill (black point)
\begin{equation}\label{eq:quatritSW12|3|4}
\mbox{spec}\left(\Delta_{(12|3|4)}\right)= 
\left\{\frac{1+3\sqrt{5}}{4}\,, \frac{1-\sqrt{5}}{4}\,, \frac{1-\sqrt{5}}{4}\,, \frac{1-\sqrt{5}}{4} \right\}\,.
\end{equation}
\end{enumerate}
\item \underline{SW kernel with two double degeneracy}:
\vspace{0.3cm}

SW kernel of  1|23|4 type:   
\hfill (purple point)
\begin{equation}\label{eq:quatritSW1|23|4}
\mbox{spec}\left(\Delta_{(1|23|4)}\right)= 
\left\{\frac{1+\sqrt{15}}{4}\,,\frac{1+\sqrt{15}}{4}\,, \frac{1-\sqrt{15}}{4}\,, \frac{1-\sqrt{15}}{4} \right\}\,.
\end{equation}
\end{enumerate}
\bigskip

All the above categories of SW kernels of a quatrit are depicted in Fig.\ref{Fig:QbQbKernelSimplex}.
The interior of a curvilinear triangle $ABC$ on $(\nu_1, \nu_2)$\--plane corresponds to the regular SW kernels. The boundary lines of the domain 
describe the double degeneracy cases:
\begin{enumerate}
\item[(a)] SW kernel of type $12|34$\---side $AB/\{A,B\}$ 
(green in color) 
with both end points $A$ and $B$ excluded: 
$$
AB/\{A,B\}:\ 
\nu_2=\frac{1}{2}-\nu_1-\frac{1}{2}\sqrt{7+4\nu_1-8\nu_1^2}\,, \qquad 
\nu_1 \in \left(\frac{1-\sqrt{5}}{4},\frac{1+\sqrt{5}}{4}\right)\,;
$$
\item[(b)] SW kernel of type $1|234$\---side $CB/\{C,B\}$ 
(red in color) 
without end points: 
$$
CB/\{C,B\}:\ \nu_2=\frac{1}{3}-\frac{1}{3}\nu_1-\frac{1}{3}\sqrt{22+4\nu_1-8\nu_1^2}\,, \qquad
\nu_1 \in \left(\frac{1-\sqrt{15}}{4},\frac{1+\sqrt{5}}{4}\right)\,;
$$
\item[(c)] SW kernel of type $123|4$\---side  $AC/\{A,C\}$ 
(blue in color) 
without end points: 
$$
\nu_2=\nu_1\,, \quad \nu_1 \in \left(\frac{1-\sqrt{15}}{4},
\frac{1-\sqrt{5}}{4}\right)\,.
$$
\end{enumerate}
The vertexes $A$ and $B$ describe a quatrit kernels with a triple degeneracy:
\begin{enumerate}
\item[(a)] SW kernel of 12|3|4 type \--- point A: 
$\nu_1=\frac{1-\sqrt{5}}{4}\,, \nu_2=\frac{1-\sqrt{5}}{4}$\,;
\item[(b)] SW kernel of 1|2|34 type  \--- point B:
$\nu_1=\frac{1+\sqrt{5}}{4}\,, \nu_2=\frac{1-3 \sqrt{5}}{4}$\,,
\end{enumerate}
while the vertex $C$ corresponds to a quatrit kernel with two double degeneracy of  1|23|4 type: $\nu_1=\nu_2=\frac{1-\sqrt{15}}{4}\,$.


%
\begin{figure}[ht]
\includegraphics[width=0.8\linewidth]{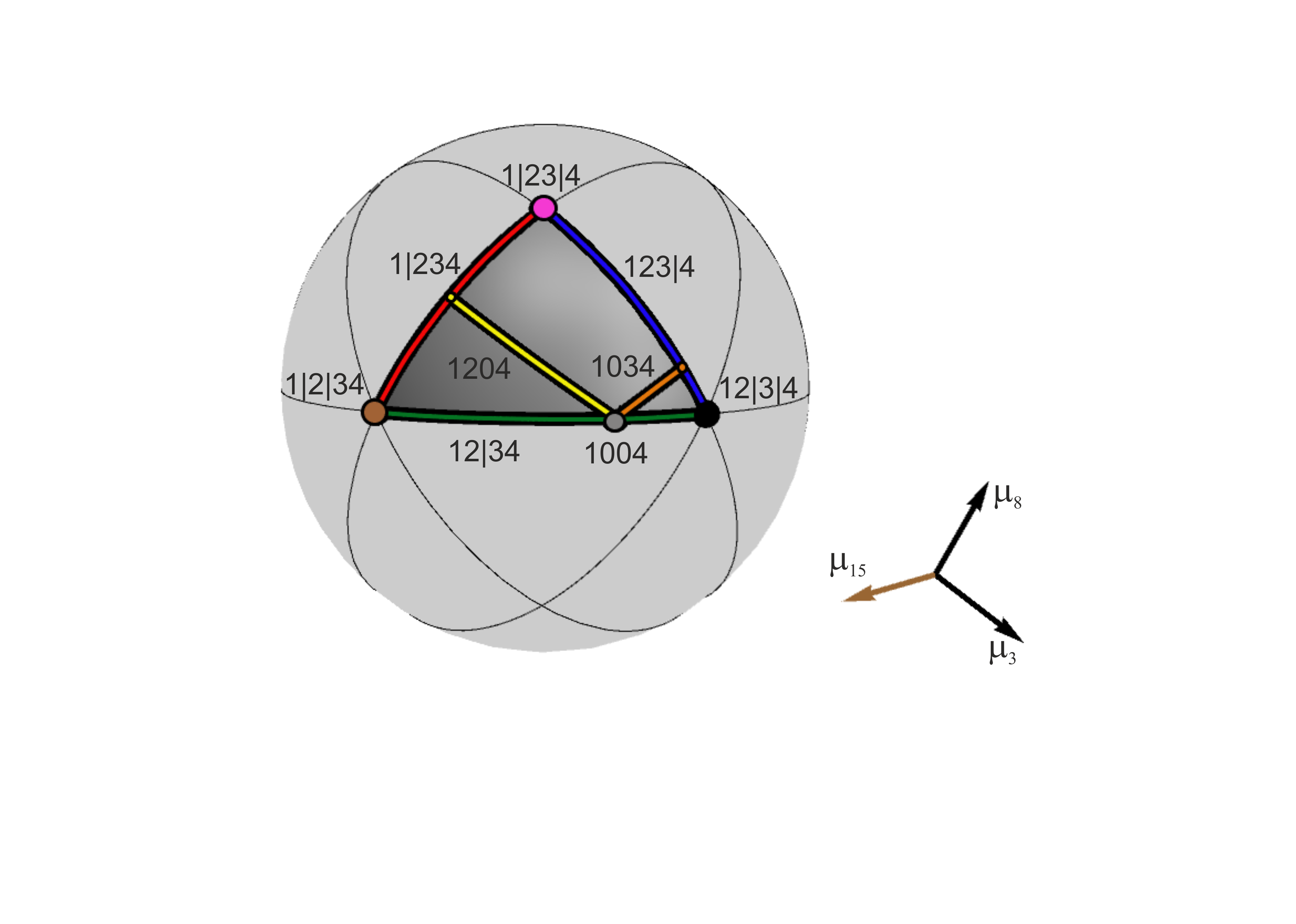}
\caption{A quatrit moduli space represented by the M{\"o}bius  spherical triangle $(2,3,3)$ on a unit sphere.}
\label{Fig:MobiusTriangle} 
\end{figure}


\noindent{$\bullet$ {\bf The singular SW kernels of a quatrit }$\bullet$} 
Among the above described SW kernels one can distinguish a set of special elements with a vanishing determinant.  These singular quatrit kernels of are listed  
below in the accordance with the increasing singularity of the determinant: 
\begin{itemize}
\item SW kernels with a simple root of the determinant:
\begin{enumerate}
\item[(a)] 1-parameter family of $1204$ type\,, $\frac{1}{3} \left(1-\sqrt{22}\right)\leq \nu <\frac{1}{2} \left(1-\sqrt{7}\right)$\,,
\begin{equation}\label{eq:quatritSWSing1204}
\mbox{spec}\left(\Delta_{(1204)}\right)= 
\left\{\frac{1-\nu+\sqrt{7+2\nu-3\nu^2}}{2}\,,\frac{1-\nu-\sqrt{7+2\nu-3\nu^2}}{2}\,,0\,,\nu \right\}\,,
\end{equation}
\item[(b)]  1-parameter family of $1034$ type\,, $\frac{1}{6} \left(2-\sqrt{22}\right)\leq \nu < 0$\,,
\begin{equation}\label{eq:quatritSWSing1034}
\mbox{spec}\left(\Delta_{(1034)}\right)= 
\left\{\frac{1-\nu+\sqrt{7+2\nu-3\nu^2}}{2}\,, 0\,, \nu\,, \frac{1-\nu-\sqrt{7+2\nu-3\nu^2}}{2}\right\},
\end{equation}
\end{enumerate}    
\item SW kernel with double zero of determinant:  
\begin{equation}\label{eq:quatritSWSing1004}
\mbox{spec}\left(\Delta_{(1004)}\right)= 
\left\{\frac{1+\sqrt{7}}{2}\,, 0\,, 0\,,\frac{1-\sqrt{7}}{2}\right\}\,.
\end{equation}
\end{itemize}

\begin{figure}[ht]
\centering
\includegraphics[width=0.6\textwidth]{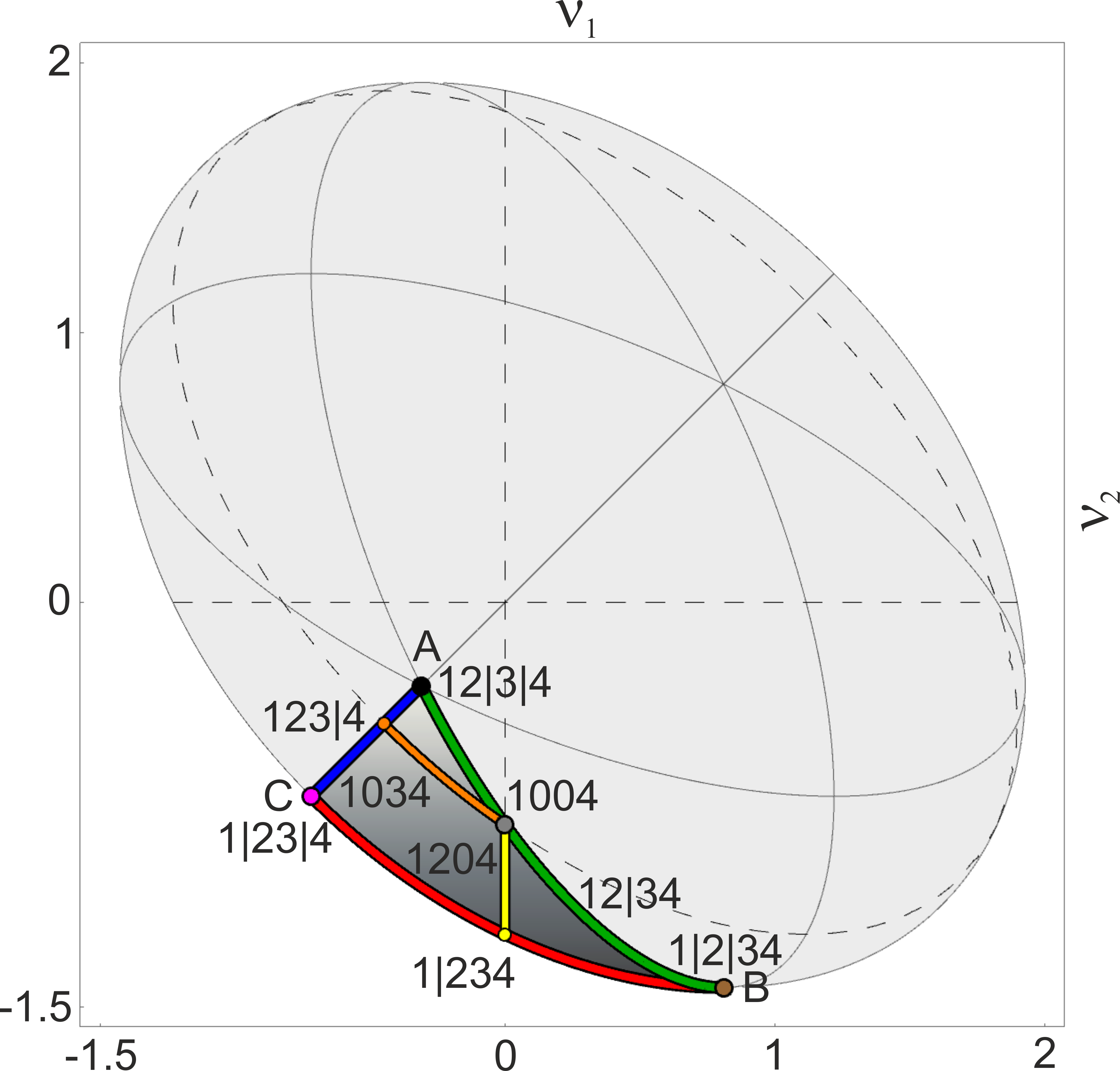}
\caption{
Mapping of the tiling of $\mathbb{S}_{2}(1)$  sphere by the M{\"o}bius  triangles $(2,3,3)$ onto a subset of the plane $(\nu_1,\nu_{2})$. 
The dashed lines represent the degeneracies of the spectrum.}
{\label{Fig:ProjectionQuatrit}}
\end{figure}

\noindent{$\bullet$ {\bf A quatrit moduli space as the M{\"o}bius  spherical triangle} $\bullet$} As it was mentioned before, the spectrum of $\Delta^{(4)}(\nu_1,\nu_2)$ is in correspondence with points on a unit 2-sphere associated with expansion coefficients $\mu_3\,,\mu_8\,$ and  $\mu_{15}$:  
$$\mu_3^2(\boldsymbol{\nu})+\mu_8^2(\boldsymbol{\nu})+\mu_{15}^2(\boldsymbol{\nu}) =1\,,
$$
which satisfy the inequalities: 
\begin{equation}\label{eq:quatritIN}
\nonumber \mu_3\geq 0\,, \quad 
\mu_8\geq \frac{\mu_3}{\sqrt{3}}\,, \quad 
\mu_{15}\geq \frac{\mu_8}{\sqrt{2}}\,.
\end{equation}
Geometrically these constraints   define  one out of 24 possible  spherical triangles  with angles $(\pi/2\,, {\pi}/{3}\,, {\pi}/{3})$ that tessellate a unit sphere.
Repeated reflections in the sides of the triangles will tile a sphere exactly once. In accordance with Girard's theorem, a spherical excess of a triangle determines a solid angle: $\pi/2 + \pi/3 + \pi/3 - \pi = 4\pi/24\,.$  
Relation between 
``flat''  representation of a quatrit moduli space  (Fig.\ref{Fig:QbQbKernelSimplex}) and its spherical realization (Fig.\ref{Fig:MobiusTriangle}) is demonstrated  by the projection pattern in Fig. \ref{Fig:ProjectionQuatrit}. 

\bigskip

\section*{Concluding remark}

The master equations (\ref{eq:master}) for kernels of the  Wigner functions  
determine the first and second degrees polynomial $SU(N)$ invariants of 
$N\--$dimensional system.
The remaining $N-2$ algebraically independent invariants parameterize  
the moduli space of SW kernels. In the  present article 
we establish relation between this moduli space and the  
orbit space of $SU(N)$ group.
Next important issue is to clarify the role these unitary invariant moduli 
parameters play in dynamics of classical and quantum systems. 
With this aim in the forthcoming publication, a detailed  analysis of the 
Kirillov-Kostant-Souriau symplectic 2-form
for the whole family of the Wigner functions will be given.

\appendix

\section{Parametrization of the moduli space 
$\mathcal{P}_N({\boldsymbol{\nu}})$}
\label{AppendixA}

As it was mentioned in the main text, the Stratonovich-Weyl  kernel can be parameterized by $N-2$ spherical angles. Each member of the Wigner functions family  can be associated with a point of subspace $\mathcal{P}_N({\boldsymbol{\nu}}) \subset \mathbb{S}_{N-2}(1)$, which is determined by the ordering of the eigenvalues of the Stratonovich-Weyl kernel.  
In order to define the $\mathcal{P}_N({\boldsymbol{\nu}})$
corresponding to the descending  ordering and by means of using kernel decomposition in Gell-Mann bases, let us represent the spectrum of the Stratonovich-Weyl kernel in the following form:
\begin{eqnarray}
\nonumber &\pi_{1}&=\frac{1}{N}\left(1+\sqrt{2}\, \kappa\sum_{s=2}^{N}\frac{\mu_{s^2-1}}{\sqrt{s\left(s-1\right)}}\right)\,,
\\
\nonumber&\vdots&
\\
\nonumber &\pi_{i}&=\frac{1}{N}\left(1+\sqrt{2}\, \kappa\sum_{s=i+1}^{N}\frac{\mu_{s^2-1}}{\sqrt{s\left(s-1\right)}}-\kappa \sqrt{\frac{2\left(i-1\right)}{i}}\mu_{i^{2}-1}\right)\,,
\\
\nonumber &\vdots&
\\
\nonumber&\pi_{N}&=\frac{1}{N}\left(1-\frac{N^{2}-1}{\sqrt{N+1}}\mu_{N^{2}-1}\right)\,.
\end{eqnarray}

Introducing the conventional parametrization for a unit sphere $\mathbb{S}_{N-2}(1)$ in terms of spherical $N-2$ angles: 
\begin{equation}
\begin{array}{lll}
&\mu_{3}=\sin{\psi_{1}}\cdots\sin{\psi_{N-2}}\,,
\\
&\mu_{8}=\sin{\psi_{1}}\cdots\sin{\psi_{N-3}}\cos{\psi_{N-2}}\,,
\\
&\vdots&
\\
&\mu_{i^{2}-1}=\sin{\psi_{1}}\cdots\sin{\psi_{N-i}}\cos{\psi_{N-i+1}}\,,
\\
&\vdots&
\\
&\mu_{N^{2}-1}=\cos{\psi_{1}}\,,
\\&&\\
&\text{with }\quad \psi_{i}\in\left[0,\pi\right],\; i=\overline{1\,, N-3} \quad \text{and }\quad \psi_{N-2}\in\left[0,2\pi\right)\,,
\end{array}
\end{equation}
and demanding the descending order of the eigenvalues, we obtain the following constraints on  the coefficients $\mu_{i}$:
\begin{eqnarray}\label{eq:ineqmu1}
&&\mu_{3}\geq0\,,\\
&&\mu_{\left(i+1\right)^{2}-1}\geq\sqrt{\frac{i-1}{i+1}}\,\mu_{i^{2}-1}\,,\quad i=\overline{2\,,N-1}\,.
\label{eq:ineqmu2}
\end{eqnarray}

Let us introduce the following notations:

\begin{equation}
\begin{array}{lcll}
&\mathcal{P}_{1}&=&\big\{\psi_{1}=0\big\}\,,\\&&&\\
&\mathcal{P}_{2}^{(k)}&=&\begin{cases}
\sin{\psi_{N-k}}=0\\
\sin{\psi_{N-(k+1)}}\cos{\psi_{N-k}}>0 
\\
\cot{\psi_{N-i}}\geq\sqrt{\frac{i-1}{i+1}}\,\cos{\psi_{N-i+1}}\,\\
0<\psi_{i-k}<\pi\,,\quad  i=\overline{k+1\,, N-1}\,,
\end{cases}
\end{array}
\end{equation}
\begin{equation}
\begin{array}{lcll}
&\mathcal{P}_{3}&=& \begin{cases}
\sin{\psi_{N-2}}>0\\
\cos{\psi_{N-2}}\geq\frac{1}{\sqrt{3}}\sin{\psi_{N-2}}\\
\cot{\psi_{N-i}}\geq\sqrt{\frac{i-1}{i+1}}\,\cos{\psi_{N-i+1}}\,\\
0<\psi_{i-2}<\pi\,,\quad i=\overline{3\,,N-1}\\\
0<\psi_{N-2}<2\pi\,.
\end{cases}
\end{array}
\end{equation}

In the introduced notations substitution of  expressions for $\mu_i$ in terms of the spherical angles $\psi_{i}$ into (\ref{eq:ineqmu1}) and (\ref{eq:ineqmu2}) shows: if $k=2\,,\cdots\,N-2$ is the biggest  natural number for which $\sin{\psi_{N-k}}=0$\,, if there is any, then the simplex is described by the restrictions $\mathcal{P}_{2}^{(k)}\subset\mathbb{S}_{N-\left(k+1\right)}(1)$ (these are some of ($N-\left(k+1\right)$)\--dimensional boundaries of the simplex); otherwise, if there is no such $k$\,, then the restrictions are $\mathcal{P}_{3}$\,. Hence, the simplex will be completely defined by
\begin{equation}
\mathcal{P}= \mathcal{P}_{1}\cup \left(\bigcup_{k=2}^{N-2}\mathcal{P}_{2}^{\left(k\right)}\right)\cup\mathcal{P}_{3}\,.
\end{equation}

Partially reducing the set of inequalities for $\mathcal{P}_{3}$\,, we get:
\begin{equation}
\mathcal{P}_{3}=\begin{cases}
0<\psi_{N-2}\leq\frac{\pi}{3}\\
0<\psi_{i-2}<\pi\,, \quad i=\overline{3\,,N-1}\\
\cot{\psi_{N-i}}\geq\sqrt{\frac{i-1}{i+1}}\cos{\psi_{N-i+1}}\,.
\end{cases}
\end{equation}

\end{document}